A Quantum of History

Essay Review of 'Quantum Theory at the Crossroads' by G. Bacciagaluppi and A. Valentini

In 1952, David Bohm demonstrated in two classic papers that a spacetime model of individual deterministic processes could consistently underpin the statistical quantum description. There was a precursor. Bohm's proposal was a partial rediscovery of Louis de Broglie's 'pilot-wave' theory of 1927. Physicists know of de Broglie primarily as the originator of the notion that material systems may display wave-like characteristics. It is not widely known that in elaborating this idea de Broglie did not replace the classical particle by a wave but rather conceived of a physical system as comprising both aspects. In one realization of this idea, he proposed that the material point is guided, or piloted, by the wave (mathematically described by the quantum-mechanical wavefunction) according to a well-defined law derived from classical Hamilton-Jacobi theory, and distributed over an ensemble according to Born's formula. De Broglie presented these ideas at the fifth Solvay conference held in Brussels in 1927. Following the conference he dropped his pilot-wave theory, taking it up again only 25 years later after becoming aware of Bohm's work. While de Broglie has priority in suggesting this model, he did not develop it far enough to prove that it could be consistently applied to generic quantum processes; it was Bohm who took this decisive step. Actually, de Broglie regarded the pilot-wave approach as a provisional point of view necessitated by the mathematical difficulties in realizing his true programme, the theory of the 'double-solution' (in which the corpuscular aspect would be incorporated as an aspect of a field introduced in addition to the wavefunction, a still unfulfilled programme).

    The fifth Solvay conference has often been presented as a key event in the history of quantum theory. Its participants included many of the giants of the day – Einstein, Schrödinger, Bohr, Born, Heisenberg, Pauli and de Broglie. It has been remembered mainly for the exchange between Bohr and Einstein, a key stage in an apparently inexorable process that resulted in the eventual triumph of the 'Copenhagen' interpretation and the ostracism of Einstein and the few others who sought a rational alternative. In their book, Bacciagaluppi and Valentini (BV) want to call attention to facets of the conference that they believe are of great importance yet have been unfairly or misleadingly reported by subsequent commentators. As a contribution to this reassessment, they have translated the proceedings (published originally in French in 1928) - comprising a series of formal talks followed by discussions - into English, and written a commentary on them. BV's principal concern is to right several historical wrongs they believe have been done to de Broglie in connection with his role in the conference, and this is the focus of this review. They argue that de Broglie's theory was one of three 'sharply conflicting' points of view presented at the meeting (matrix mechanics, wave mechanics and pilot-wave theory) that were 'on an equal footing'. They seem to suggest, as hinted at by the title, that there was uncertainty as to which view might be crowned the official creed of theoretical physics. At first sight this seems a somewhat fanciful notion which, if true, should instigate a major revision of the history of the period when the official interpretation of quantum mechanics was being cemented. The authors hope that their study will prompt a reexamination of current approaches to interpretation but are somewhat vague about how a detailed investigation of this remote singular episode might contribute to that goal.

Among the claims that BV wish to contest that they assert have been made in the fifth Solvay-related literature are: that a consensus was reached at the conference in favour of the interpretation of Bohr and Heisenberg (BV: no agreement was reached as to which view was dominant); that the pilot-wave theory received little attention (BV: it was discussed extensively); that de Broglie presented his theory only for a one-body system (BV: he presented the many-body case); that de Broglie abandoned his approach because he could not answer a criticism by Pauli regarding the theory's ability to treat scattering processes (BV: he essentially gave the correct answer to Pauli at the conference); and that after the conference de Broglie adopted the views of Bohr and Heisenberg (BV: he continued to express doubts). They write as if there is considerable novelty in their observations.

It is the case that some writers have been sloppy in reporting the congress and its aftermath. It is true, for example, that de Broglie did present his theory for the many-body case (although he deemed the configuration space wavefunction to be 'fictitious', and he did not apply it to any problems), and that he gave it up for reasons other than Pauli's criticism (see below). Beyond these minor corrections of the record, however, BV's claims, their implied claim to novelty, and their supporting arguments, need to be assessed carefully in the light of the full historical record. Also, it is important to appreciate that, even if all the authors' claims were granted, a far greater case could be made; the de Broglie-Bohm theory has been wilfully misrepresented or ignored throughout its history and that story is still to be written.

Among BV's guilty commentators is de Broglie. Their attitude to him is ambivalent; they simultaneously eulogise him for his 'new dynamics' while rejecting his version of events when it is at variance with theirs. De Broglie is a major source of information about the fifth Solvay conference and indeed much of the key material discussed by BV is recounted concisely in his books (e.g., [1] chap. 10; [2] chap. 8; [3] chaps. 8 and 14). A problem for BV is that de Broglie's accounts differ from theirs in respect of details that they regard as important – in particular, in their assertions that the theory received much attention, that de Broglie essentially answered Pauli already, and that he did not subsequently defer to Bohr and Heisenberg. They deal with this inconvenience by dismissing de Broglie's reports as 'mistaken 'recollections'…decades later' which 'do not seem a reliable guide to what actually happened circa 1927' and 'are belied by the published proceedings'. In fact, the many accounts de Broglie gave of his experience are remarkably consistent, and they did not commence 'decades later' (not that BV offer any reason why one should automatically distrust the later accounts; indeed, they treat de Broglie's 1963 recollections about the genesis of his ideas in 1923 as unproblematic). One finds a fairly complete and honest description in an essay published by de Broglie in 1941 ([2], chap. 8), only 14 years after the conference (this is not cited by the authors - they refer only to Cushing's report of it, and Cushing gave few details and referred only to the 1947 French reprint). To give a flavour, it's worth reproducing the same story extracted from the more detailed account de Broglie published in 1956 ([3] chap. 14):

'At the Solvay Congress…I balked at the difficulties of mathematically justifying the double-solution point of view, and I contented myself with a presentation of the pilot-wave point of view…while a few of the "old guard" (Lorentz, Einstein, Langevin, Schrödinger) insisted on the necessity of finding a causal interpretation of Wave Mechanics - without, however, coming out in favour of my efforts - Bohr and Born, along with their young disciples (Heisenberg, Dirac, etc.), came out categorically in favor of the new purely probabilistic interpretation that they had developed, and they

did not even discuss my point of view. Pauli was the only one to present a definite objection to my theory…'

De Broglie then gives what BV state is 'the first proper analysis of Pauli's example in terms of pilot-wave theory' (which statement tacitly concedes that his reply in 1927 was inadequate). He goes on:

'In the months that followed the Solvay Congress of October 1927,
I abandoned the pilot-wave approach that I had maintained. But not because of Pauli's objection. For, as I say, I thought I had found the way in which to overcome it. Rather, I abandoned the pilot-wave approach for other, more general, reasons…'

De Broglie then sets out the reasons he gave in 1930 [4] (to do with how a probability wave may be ascribed physical reality). He concludes:

'Such were the considerations that led me, in 1928, to abandon the pilot-wave theory as untenable… So I joined the ranks of the purely probabilistic interpretation put forward by Born, Bohr and Heisenberg…To my knowledge, from 1928 to 1951 no serious attempt was made to effectively construct any other interpretation more closely approaching classical conceptions.'

So, according to de Broglie himself, his theory received little meaningful attention at the conference, he abandoned it soon after (and not because of Pauli), and he henceforth adopted the views of Bohr and Heisenberg. Do the proceedings, or perhaps other sources, cause us to doubt the veracity of de Broglie's version?
    The conference discussions on the pilot-wave theory are certainly of interest. Particularly notable, given the novelty of the theory, is the sharpness of the commentary: Born and Pauli each raised questions about how the theory would handle scattering processes, Pauli noted the key connection between conservation laws and trajectories, Schrödinger noted the possibility of alternative pilot-wave theories and observed that generic atomic motions would not be circular, Ehrenfest asked whether an atomic electron could be at rest in a stationary state, Einstein asked whether photons could have zero speed, and Lorentz pointed out that de Broglie's variable mass could be imaginary (an observation that effectively sank the Klein-Gordon equation-based theory de Broglie was using to explain the pilot-wave idea; de Broglie did not reply). The most significant contribution was unquestionably Pauli's scattering problem and subsequent accounts have quite reasonably concentrated on that. But, in the end, the contributions from the attendees other than de Broglie and his circle, although pertinent, often read like the polite remarks of interested bystanders; one would read a lot into these exchanges to conclude that the attention the theory received was of any real significance. There was no genuine dialogue that might inspire further study. Even Einstein's statement that in addition to the Schrödinger wave one should localize 'the particle during…propagation. I think that Mr. de Broglie is right to search in this direction…' was not an endorsement of de Broglie's specific pilot-wave model (he later described Bohm's theory as 'too cheap'). De Broglie answered some of the questions but, in truth, he was ill equipped to offer a trenchant response and his replies are at times unconvincing (e.g., his response to a query by Kramers about how radiation pressure is exerted). One of his difficulties, with which BV do not fully engage, is that he was working with an untenable relativistic theory which entailed a 'probability' density of indefinite sign

and material trajectories that pass through the light cone. He did present a (consistent) non-relativistic version of the theory but in the one-body case this was presented as a limiting case of the (inconsistent) relativistic treatment. As noted above, the guiding agent, Schrödinger's wavefunction, he regarded as fictitious. It was all a bit of a muddle. He had not developed the theory far enough to see how to apply it in sufficient generality, and he had not thought through his presentational tactics (in his memoir of 1941 ([1] chap. 10) he lamented that adopting the pilot-wave reasoning 'had greatly weakened my position'). In reply to the Pauli scattering objection he 'could half-see a possible answer, but without being able to make it completely precise' ([1] chap. 10) which indeed seems an accurate summary of his brief reply in the proceedings. In contrast to BV's reading, one gains an inkling of how the conference contributed to the demise of de Broglie's conviction in his theory and it is not clear why BV wish to contest this. In fact, it is not noted by BV that de Broglie had already expressed doubts about the pilot-wave concept on the eve of the conference, to do with its compatibility with Born's interpretation of superpositions ([2] chap. 8). BV observe that a significant fraction of the text is devoted to consideration of de Broglie's ideas but this does not signify that his approach had a proportionate impact. It is after all an historical fact that research in this field died at this point. BV's claim that de Broglie did not immediately promote Bohr & Co is based on some ambivalent incidental remarks he made (not in the proceedings) and this evidence is really too meager to counter de Broglie's own version. De Broglie has comprehensively explained what happened. What is the problem?

    A methodological problem with BV's analysis is that they offer an almost literal reading of the proceedings with no allowance for what may lurk in the margins. After all, as BV themselves point out, the Bohr-Einstein dialogue, for which the congress is renowned, does not feature in the proceedings at all! When de Broglie says ([3] chap. 8) '…the purely probabilistic interpretation…was very clearly the one preferred by most of the scientists present…The objections raised against my approach, as well as the almost unanimous acclaim of the members of the Congress for the Bohr-Heisenberg interpretation…made a very great impression upon me', does this not suggest that there is a more subtle reading of the transcripts? While there was no definitive statement in the proceedings announcing that Bohr and Heisenberg had 'won', the desire to close the subject, to regard the principles of quantum mechanics as essentially already complete and not further modifiable, was manifest. And the belief that it is illusory to attribute simultaneously well-defined position and velocity variables to an electron was already widely accepted by 1927, as de Broglie noted in his talk. A new type of utilitarian knowledge was being made; not even Einstein could stand in its way. Born and Heisenberg were offering novel algorithms to solve practical problems, with Bohr supplying an exotic epistemological veneer that appeared to prove (even though no one seemed to understand why) that this was all the physicist could aspire to because there is in principle no way of imagining the meaning of the theory. Against this, de Broglie was presenting visual representations harking back to 'classical conceptions' (using a theoretical framework that he privately regarded as inferior); these people were talking past each other. The conflicting points of view were hardly 'on an equal footing'. The outcome was that quantum mechanics became a theory of formal splendour but conceptual impoverishment and from thereon the subject acquired the air of paradox and mystery that dominates its discourse to this day. One has to question whether this fundamental transformation of theoretical physics can be comprehended on the basis of the historical record of a single event plucked from its social environment.

The authors' evangelical style leads them to adopt peculiar and unnecessary stances, particularly in the way Bohm's 1952 work is treated. In an apparent attempt to establish the superiority of de Broglie's approach they set up a fictitious distinction with Bohm, rigidly associating de Broglie with a 'first-order' law of motion involving velocity and Bohm with a 'second-order' law based on acceleration. They claim that 'the extent to which de Broglie's dynamics departs from classical ideas was unfortunately obscured by Bohm's presentation of it…in terms of…a…quantum potential, a formulation that…seems artificial and inelegant compared with de Broglie's'. It is amusing to note that it was de Broglie (along with Madelung) who introduced the 'quantum potential', and indeed gave it that name (e.g., [4])). While de Broglie did present his theory in first-order terms, he also derived the second-order version and the authors cite no evidence that he disparaged the notion of force in this context. On the contrary, in the proceedings de Broglie invokes the 'force of a new kind' as a causal agent that will 'curve the trajectory' in regions of interference, an observation that BV treat not as an important insight deserving analysis but as incidental. De Broglie made extensive use of the quantum potential and quantum force in his later work (e.g., [3]). The distinction with Bohm is created artificially by selective reporting. Bohm clearly stated the first-order equation for calculating trajectories in his first 1952 paper ([5] sec. 4). He also worked with the force-based version but it is incorrect to claim that for him the velocity constraint entered only as a supplementary initial condition on the Newton-type law. Indeed, in a 1962 paper (reprinted in [6] chap. 4) Bohm presented a purely first-order version of his theory. So, de Broglie and Bohm both used the first- and second-order formulations of the pilot-wave theory, depending on the context and the issue at hand. And it is perfectly natural that they should do so since such interplay between laws is an obvious feature of classical mechanics in its Hamilton-Jacobi formulation to which they were both appealing. Once one is in possession of the Hamilton-Jacobi function, one of Jacobi's equations combined with one of Hamilton's equations supplies the required first-order law to compute the trajectories. One can make either the first- or second-order versions look more significant just by adopting a fragmentary approach to dynamics. It was not de Broglie but more recent practitioners who have fetishized the first-order formulation and neutered the potent conceptual legacy of the pilot-wave pioneers. At the present stage in the development of trajectory theories a progressive outlook would recognize that embracing the widest possible portfolio of concepts is more likely to be productive than promoting one partial view over another as a means of what is often no more than marking out academic territory. In this connection it is noted that, far from obscuring the relation between classical and quantum behaviour, it is precisely the quantum potential that enables one to gauge how 'quantum' or 'classical' a system is, and the quantum force is crucial to a full account [7]. Thus, in this context, the guidance laws of *both* orders must be applied. In any case, as mentioned above, de Broglie did not attach to the pilot-wave theory the fundamental significance that BV do since for him it maintained an unacceptable dualism that he sought to transcend using the double-solution theory.

There *are* significant differences between de Broglie and Bohm. Bohm had a deeper understanding of the causal theory and he used its insights to develop a series of profound philosophical analyses of modern physics. While Bohm was prepared to embrace the theory's radical implications, such as nonlocality, de Broglie was more timid and there is evidence in his writings that he coveted a return to 'classical conceptions' (cf. the quote above). And de Broglie's concept of pilot wave was *not* Bohm's or the one customarily presented nowadays for he maintained the view that it is untenable to attribute reality to the probabilistically interpreted wavefunction ([3] chap. 15).

For physicists interested in a key segment of quantum history where unconventional but valuable ideas were discarded, the transcripts of the fifth Solvay conference are enlightening and recommended reading. Notwithstanding the reservations expressed above, this book could play a role in guiding readers to the original works of de Broglie and Bohm, and perhaps promote a more open-minded appreciation of their contributions than is customary even now.

Peter Holland
Green Templeton College
University of Oxford
peter.holland@gtc.ox.ac.uk